\documentclass[pra,nofootinbib,onecolumn]{revtex4}

\usepackage{graphicx}
\usepackage{amssymb}
\usepackage{amsmath}
\usepackage{amsfonts}
\usepackage{epstopdf}
\usepackage{epsfig}
\usepackage{wrapfig}

\begin{document}

\title{A possible statistical mechanism of anomalous neutrino velocity in OPERA experiment?}
\author{Robert Alicki}
\affiliation{Institute of Theoretical Physics and Astrophysics University of
Gd\'ansk, Poland }

\begin{abstract}
The set of kinetic equations describing the process of conversion of a beam of protons into mesons and then to neutrinos is solved. The asymptotic evolution of the density profile of neutrinos is essentially the same as that obtained in the  previous version of the note for a simple model of uniformly damped wave-packet. It shows again that the recently reported "superluminal neutrinos" \cite{OP} could be considered as a purely statistical effect due to the fact that the detected neutrinos represent a biased sample of initial protons.
\end{abstract}

\maketitle

Consider a one-dimensional  model which describes propagation of a beam of particles with the speed of light. The beam consists of protons with the density $p(x,t)$, mesons with the density $m(x,t)$ and neutrinos with the density $n(x,t)$. The kinetic equations read
\begin{eqnarray}
\Bigl(\frac{\partial}{\partial t} + c\frac{\partial}{\partial x} + \Gamma_p \Bigr)p(x,t) &=& 0 \nonumber\\
\Bigl(\frac{\partial}{\partial t} + c\frac{\partial}{\partial x} + \Gamma_m \Bigr)m(x,t)&=& \gamma_p p(x,t) , \label{kinetic}\\
\Bigl(\frac{\partial}{\partial t} + c\frac{\partial}{\partial x} + \Gamma_n \Bigr)n(x,t) &=& \gamma_m m(x,t)\nonumber
\end{eqnarray}
where $\Gamma_p ,\Gamma_m ,\Gamma_n$ are damping rates for protons, mesons and neutrinos respectively, while $\gamma_p$ and $\gamma_m$ describe transformation rated of protons into mesons and mesons into neutrinos.
\par
The system of equations (\ref{kinetic}) can be easily solved and for the initial conditions : $ p(x,0)= f(x)$ , $ m(x,0)= 0$ and $ n(x,0)= 0$ yields  
\begin{eqnarray}
n(x,t) &=& \frac{\gamma_p \gamma_m}{\Gamma_p -\Gamma_m} \Bigl( \frac{1}{\Gamma_m - \Gamma_n} - \frac{1}{\Gamma_p - \Gamma_n}\Bigr) e^{-\Gamma_n t}f(x - ct)\nonumber\\
&+&\frac{\gamma_p \gamma_m}{\Gamma_p -\Gamma_m} \Bigl( \frac{1}{\Gamma_p - \Gamma_n}e^{-\Gamma_p t} - \frac{1}{\Gamma_m - \Gamma_n}e^{-\Gamma_m t}\Bigr) f(x - ct), \label{sol}
\end{eqnarray}
Under the physically justified condition $\Gamma_n << \Gamma_p , \Gamma_m$ and for $t>> 1/\Gamma_n$  the expression (\ref{sol}) simplifies to
\begin{equation}
n(x,t) \simeq \frac{\gamma_p \gamma_m}{\Gamma_p \Gamma_m}e^{-\Gamma_n t}f(x - ct) .
\label{simple}
\end{equation}
which essentially coincides with the result of the previous version of this note. For the initial Gaussian density profile (concentrated at the origin for $t=0$) the neutrino density evolves asymptotically as
\begin{equation}
n(x,t) = N \exp\Bigl\{-\frac{1}{2d^2}[x - ct]^2\Bigr\} \exp\{- \Gamma_n t\} .
\label{Gaussian}
\end{equation}
where $d$ is the width of the density profile and $N$ - the normalization constant. The time resolved  density profile (\ref{Gaussian}) at the point $x = L >> c/\Gamma_n$ can be recast into the form
\begin{equation}
n(L,t) = N \exp\Bigl\{-\frac{\delta L}{L}\bigl(\frac{L}{d}\bigr)^2\Bigr\}\exp\Bigl\{-\frac{1}{2d^2}\bigl[(L -{\delta L}\bigr) - ct \bigr]^2\Bigr\} 
\label{Gaussian1}
\end{equation}
where the shift
\begin{equation}
\delta L = \frac{\Gamma_n d^2}{c}  
\label{shift}
\end{equation}
can be misinterpreted as a consequence of a higher velocity $v \simeq c (1 + \delta L/L)$. The origin of this shift is purely statistical, the neutrinos created at the longer distance from the proton source,  and hence arriving earlier at the detector, are slightly less damped than those created at the shorter distance.
\par
To apply this result to the OPERA experiment \cite{OP} one should notice that in the experiment one compares the density profile for the protons - the "grandparents of the neutrinos" with the  profile for neutrinos. The former is measured under conditions where damping can be neglected and the time-resolved shape does not display the shift $\delta L$. On the other hand, the highly damped density profile of neutrinos is shifted. The damping rate $\Gamma_n$ describes an overall effect of all processes which reduce the fraction of  neutrinos produced by the source in CERN which are detected in Gran Sasso Lab. Finally, one can compare the relevant numbers. The initial proton density profile is composed of five peaks which can be approximated by the Gaussians of the width $d\simeq 600 m$. This value combined with the value of $L\simeq 700 km$ and the measured relative shift $\delta L/L \simeq 2.5\times 10^{-5}$ yield the overall damping factor (the prefactor ${\gamma_p \gamma_m}/{\Gamma_p \Gamma_m}$ is neglected) 
\begin{equation}
{\cal D} \simeq  \exp\Bigl\{-\frac{\delta L}{L}\bigl(\frac{L}{d}\bigr)^2\Bigr\}\simeq 10^{-13} . 
\label{damping}
\end{equation}
This value is consistent (within a reasonable accuracy) with the ratio $N_{\nu}/N_p \sim 10^{-16}$ where $N_{\nu}\sim 10^{4}$ is the total number of  neutrinos detected in OPERA and $N_p  \sim 10^{20}$  is the number of the corresponding protons.

\end{document}